\def\year{2022}\relax
\documentclass[letterpaper]{article} 
\usepackage{aaai22}  
\usepackage{times}  
\usepackage{helvet}  
\usepackage{courier}  
\usepackage[hyphens]{url}  
\usepackage{graphicx} 
\usepackage{natbib}  
\usepackage{caption} 
\DeclareCaptionStyle{ruled}{labelfont=normalfont,labelsep=colon,strut=off} 
\usepackage{algorithm}
\usepackage{algorithmic}

\usepackage{newfloat}
\usepackage{listings}
\floatstyle{ruled}
\newfloat{listing}{tb}{lst}{}
\floatname{listing}{Listing}
\title{Modes of Analyzing Disinformation Narratives With AI/ML/Text Mining to Assist in Mitigating the Weaponization of Social Media}
\author{
    Andy Skumanich\textsuperscript{\rm 1}\equalcontrib,
    Han Kyul Kim\textsuperscript{\rm 2}\equalcontrib
}
\affiliations{
    \textsuperscript{\rm 1}	Innov8ai Inc.\\
    \textsuperscript{\rm 2}	University of Southern California\\

    DrAndySku@gmail.com, hankyulk@usc.edu \\
}

\usepackage{bibentry}

\begin{document}

\maketitle

\begin{abstract}
 This paper highlights the developing need for quantitative modes for capturing and monitoring malicious communication in social media. There has been a deliberate ``weaponization" of messaging through the use of social networks including by politically oriented entities both state sponsored and privately run. The article identifies a use of AI/ML characterization of generalized ``mal-info," a broad term which includes deliberate malicious narratives similar with hate speech, which adversely impact society. A key point of the discussion is that this mal-info will dramatically increase in volume, and it will become essential for sharable quantifying tools to provide support for human expert intervention. Despite attempts to introduce moderation on major platforms like Facebook and X/Twitter, there are now established alternative social networks that offer completely unmoderated spaces. The \textbf{paper presents} an introduction to these platforms and the initial results of a qualitative and semi-quantitative analysis of characteristic mal-info posts. The \textbf{authors perform} a rudimentary text mining function for a preliminary characterization in order to evaluate the modes for better-automated monitoring. The action examines several inflammatory terms using text analysis and, importantly, discusses the use of \textit{generative algorithms} by one political agent in particular, providing some examples of the potential risks to society. This latter is of grave concern, and monitoring tools must be established. \textbf{\textit{This paper presents}} a preliminary step to selecting relevant sources and to setting a foundation for characterizing the mal-info, which must be monitored. The AI/ML methods provide a means for semi-quantitative signature capture. The impending use of ``mal-GenAI" is presented. The\textbf{ main findings} are: (1) we introduce specific politicized social channels of note; (2) we present viable indicative AI/ML modes of characterizing the output of these channels for\textit{ capturing and tracking} mal-info; (3) we provide an approach for qualitative and semi-quantitative \textit{signatures} of mal-info; (4) we flag the impending use of mal-GenAI in this context. We emphasize that mal-info in social media will be exploited by state sponsored malicious actors, and it is essential to initiate studies to capture, track, and mitigate the disinformation. We present a representative case study mode which, when expanded upon, can help mitigate the weaponization of social media intended to attack various subgroups or communities. This is a serious challenge for the times and needs urgent attention.
\end{abstract}

\section{Introduction}

Social networks have had a profound impact on the way we communicate, share information, and interact with each other. They have become a central part of modern society, enabling people to connect with each other regardless of their location, share their thoughts and opinions, and participate in online communities \cite{osman:hal-04084263, bakshy2012role}. However, the rise of social networks has also led to a number of challenges, including the spread of deliberate disinformation \cite{10.1145/3543873.3587348}, cyberbullying \cite{GIUMETTI2022101314}, and the propagation of hate speech and extremist ideologies \cite{10.1145/3583067}, along with ``mal-info" (malicious information). Note that mal-info can be defined as \textit{genuine info that is intentionally manipulated to cause harm}, but it shares the domain with e.g. hate speech (but not specifically with dis-info). Worse still is the deliberate use of mal-info for the ``weaponization" of social media, where harm is the desired outcome. \cite{weponizagion1, weponizagion2} Additional concern is given the now-impending use of ``mal-GenAI" which has already started to be implemented. There are already multiple indications of the harmful impacts resulting from mal-info narratives, such as mass shootings and the ``swatting" of citizens.

There is already a substantial body of published work on the topic of hate speech in social media, as evidenced by representative publications in the references \cite{matamoroshate, castano2021internet}. Additionally, a recent review of papers \cite{Review2023} has addressed concerns about the weaponization of social media for hate speech. However, it's crucial to highlight that their assessment is confined to standard social media channels, and they advocate for improving best practices to study it. In this paper, we go beyond these limitations, addressing both elements and providing initial yet indicative practices that can be utilized. Other very recent papers touch on the developing issues of AI generated misinformation where they highlight the need for monitoring (a point we directly address below). \cite{AImisinformation}

We present concrete examples of mal-info signatures derived from social media channels like Gab, Gettr, and Bitchute, which lack controls compared to mainstream platforms, offering a reduced set of guidelines and principles. Their popularity among certain segments of the population is attributed to their permissive stance on various forms of misinformation. While these networks champion themselves as spaces for \textit{free expression}, they also raise questions about the role of social networks on our perceptions, attitudes, and the potential ramifications for our democratic institutions and processes. Despite these pressing issues, very few articles have delved into these emerging social networks \cite{10.1177/01634437221111943}, with the focus predominantly on platforms like X/Twitter \cite{10.1145/3583067}. Furthermore, an imminent concern lies in the potential use of AI and Generative AI (GenAI) by these actors to further propagate mal-info. This latter is of grave concern, and a broad array of tools must be employed to develop ``guard rails" to mitigate the potential harm. The issue of misinformation and mal-info has emerged as a serious issue not just for online content providers but for society at large. Within this context of providing industry-based tools, this paper develops specific applications with illustrative examples. While several industry participants already engage in veracity assessment, many rely on human examiners to evaluate the content. Our aim is to supplement these manual efforts with useful modes and algorithms that can be used for Capture, Tracking, and allow for (potential) Response (``CTR" in our parlance). The more tools and analysis at the disposal the more effective the efforts.

This paper addresses four key points related to tackling the weaponization of social media: (1) describing new social media channels of importance that are concentrated for mal-info; (2) providing viable modes of capture (3) initial indicative development of signatures for mal-info elements; and (4) elucidating the impending issue of GenAI within this mal-info context. The overarching goal is to formulate potential best practices that can be employed for Capture/Track/Respond (CTR), providing valuable assistance to human experts. While the primary focus is on Capture, this paper also lays the groundwork for Tracking and allows for a framework for Response.

In this paper, we aim to illustrate the weaponization of mal-info on social media by providing specific examples. We introduce three newer (fringe) social networks, Gettr, Bitchute, and Gab with a focus on the latter. We expand on other research which has considered the extraction of disinformation from niche social media, and observed the propagation of state sponsored propaganda into social media channels \cite{LinGovDis}. \cite{SkuKimVancouver} Importantly, we present the recent advancements driven by these platforms in utilizing generative algorithms. Finally, we touch on the societal risks posed by such algorithms. Our conclusion emphasizes the urgency for the AI community to develop rapid responses to counter these risks and underscores the importance of creating AI/ML-assisted solutions.

\section{Newly Engendered Fringe Social Networks}

New social networks were created in reaction to perceived censorship and to moderation on established social media platforms. Some users feel that their ``freedom of speech" was being limited on platforms such as X/Twitter and Facebook and that their content is being unfairly targeted or removed. This is a broader discussion beyond the scope of this paper as it is part of the determination of what constitutes \textit{free speech}. As shown by \cite{stocking2022role}, more and more Americans are using these platforms for news ($6\%$ in 2022). Using an estimate with Similarweb\footnote{https://www.similarweb.com/}, would indicate that the subscription doubled in 2023 highlighting the growing influence.  By contrast Mastodon is a decentralized social media platform but still retains moderation guidelines and content restriction.  Another platform is Koo which actively removes mal-info and disruptive content. The contrast with these latter platforms highlight the extreme nature of the sites we study below.


These newer fringe platforms often have little to no moderation and enforce lenient content policies. Although this allows for a nominally wider range of opinions and viewpoints to be shared, it provides a basis for mal-info, and hate speech, harassment \cite{abarna2022identification}, and misinformation.  The mal-info can spread unchecked. In the spectrum of fringe social networks, three emerge as frequently used  channels, namely, Gab, Gettr, and Bitchute. We consider these non-moderated networks in the paper, although it must be noted that even some of the nominally moderated social networks are questionable in their limitation of mal-info. Still the analysis we apply to the non-moderated shows the egregious examples showing the risks and concerns. The unmoderated networks amplify the mal-info with an echo chamber effect as we have qualitatively observed in reading the posts. We measured that for a particular mis-informational topic, the posting level in X/Twitter was at about half the level of posts per period as compared to Gab.

Gab is a social networking platform launched in 2016 and bills itself as an unfettered speech (so-called \textit{free speech}) alternative to mainstream social media sites. It was created in response to the perceived censorship of conservative views on traditional social media sites. Gab allows users to post messages called ``gabs," share photos, and interact with other users. It has been observed as being a platform for hate speech and far-right extremism.

Gettr is a newer social media platform that was launched in 2021 by former President Donald Trump's senior adviser, Jason Miller. It is marketed as a ``cancel-free" platform that supports unfettered speech and allows users to share their opinions without any type of moderation. Gettr's features are similar to those of Twitter, allowing users to post short messages called ``gettrs," share photos and videos and interact with other users.

Bitchute is a video-sharing platform that was launched in 2017. It was created in response to perceived censorship of fringe or provoking views on traditional video-sharing sites like YouTube. Bitchute allows users to upload, share, and view videos on various topics, including news, politics, and entertainment. It has been observed as being a platform for conspiracy theories and hate speech. 

The main difference between these platforms is their focus and features. Gab and Gettr are primarily social media platforms that allow users to share short messages and interact with other users. Bitchute, on the other hand, is a video-sharing platform that allows users to upload and view longer-form content. Additionally, Gab has been associated with far-right extremism, while Gettr is marketed as a nominally more mainstream platform. All three platforms have been observed to tolerate hate speech and conspiracy theories, and represent focal points of concern as these channels are not diminishing and must be included in analysis.

One of the central themes of our findings is that the level of mal-info generation is substantial. We observe that it can often include an element of anti-semitism for the segments we studies. In addition, we note that both state sponsored talking points and parallel commercial talking points often are in alignment and are directed to mal-info. The overall aspect is that subsets of the population can be metaphorically attacked. A disturbing dimension of this metaphoric attack is that it can become actualized into real-world events.

\section{Analysis of Gab as a Representative Example}

Other work has introduced the developing nature of Gab for promoting mal-info \cite{GabCovid}. In addition, semi-quantitative analysis with text mining showed the potential to capture and track elements of mal-info in Gab \cite{SkuKim_sss}. We expand on those preliminary research directions  to examine in further detail the fringe channels. 

In order to do a deep dive into the different narratives spread by these posts, we applied the method of keyness analysis. This approach from the fields of corpus linguistics and corpus-based discourse analysis is directed at identifying key items (e.g. words) in a target corpus in relation to a reference corpus based on the
frequencies of items in both corpora. As such, a keyness analysis can support an exploratory approach to texts that gives an indication of their aboutness. The keyness metric chosen for this paper is that of Log Ratio, which is defined as the binary log of the ratio of relative frequencies.  This gives a measure of the actual observed difference between two corpora for a key item (rather than a measure of statistical significance). The advantage of this is that it allows for the sorting of items by the size of the actual frequency difference between the corpora, enabling us to find the top \textit{N} most key items.   We take a representative set of Gab data from March 2022 to August 2022 and examine the terms, with the results as shown in Table 1.  The text mining provides some signatures of the narrative which can be correlated to concurrent events such as the mass shooting in May 2022, where the words \textit{unum}, \textit{snitch} reflect a reaction of the ultra-right community.
The results are indicative of narrative flow which, although requiring Human Intelligence, allows for interpretation and subsequent analysis. Table 1 shows the key items by month. It is interesting to note the different changes in narratives. For example, in March, users are mostly talking about Covid by using words such as \textit{wuhan} and the \textit{currentthing}. Also, they are referring to the war in Ukraine with the word \textit{quagmire}. Then, in May-June, users seem to react to the Mass Shooting in Uvalde, Texas and the political reaction in favour of gun control using terms such as withheld or unum and criticizing people betraying their ideology (snitch). Finally, in July-August, they return to more usual narratives (in their ideology) by criticizing The Establishment (shitlibs), anti-Semitic organizations such as ADL or AIPAC and other far-wing ideology like Qanon. 
\begin{table}
    \centering
    \begin{tabular}{c|c}
         Month& Keyness terms\\
        \hline
         March&  wuhan, thecurrentthing, quagmire\\
         April& regimechange, extortion, falseflags\\
         May&  withheld, outlawed, unum\\
         June& snitch, plunder, accountability\\
         July& shitlibs, thinktank, fedbois\\
         August&  qanoncuck, adl, aipac\\
    \end{tabular}
    \caption{Keyness Terms for Gab}
    \label{tab:my_label}
\end{table}

Further analysis can be done with network figures and one is shown for narratives using Network analysis with the co-occurrence of N-grams (Fedoryszak, 2019), hashtags in this case, for Moron Labe (a right-wing term) from January 1, 2022, to April 26, 2023, using Pyvis (Giancarlo et al., 2020). For the convenience of visualization, we kept only hashtags that co-occur more than 50 times. These are useful data analytic visualization modes is that they provide a size-to-extent indication. The larger the star the more the prevalence. These clusters can be used as a characterization mode again to develop trends and changes.
\begin{figure}
    \centering
    \includegraphics[width=1\linewidth]{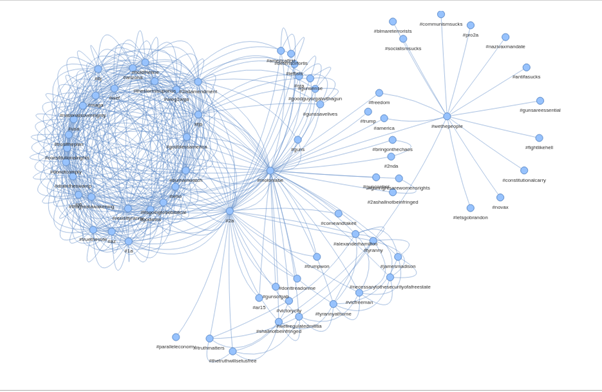}
    \caption{Example of a Network Graph for key Right-wing term of Moron Labe}
    \label{fig:enter-label}
\end{figure}
This type of signature is illustrated in Figure 1, and is indicative of modes of characterization.  For this paper we present only a brief view of the possible narrative capture signatures but which can allow for both qualitative and quantitative analysis.  Here we focus on the concern of Generative AI in this domain as follows.

An additional critical dimension is not just the sharing of mal-info and deliberate disinformation, but also the developing concern of Generative AI to amplify the harmful messaging. To demonstrate the impact of this fringe social network and its generative AI capabilities on the spread of mal-info and, in particular, hate speech (as a stand-in for weaponizing speech), we conducted exploratory text mining on interactions between Gab users and one of Gab's generative AI bots, \textit{Conspiracy AI}. In December 2023, Gab introduced a series of these AI-based chatbots, each designed with unique biases and worldviews as part of an initiative to ``embrace and diversify bias" \footnote{https://news.gab.com/2024/01/gabs-vision-for-2024-an-uncensored-ai-platform/} \textit{Conspiracy AI}, among the array of chatbots on Gab, is specifically tailored to share information about conspiracy theories. Users on Gab can easily tag Conspiracy AI in their messages, prompting the bot to generate responses related to conspiracy theories. 

For the purposes of this paper, as no API is available for Gab, we developed a custom scraper utilizing Python's Selenium library \cite{10.5555/2655462}. This tailor-made scraper was employed to crawl 119 prompts initiated by 36 users, along with their corresponding responses from \textit{Conspiracy AI}. The dataset covers interactions from the initial prompt made to Conspiracy AI on December 20, 2023, to December 31, 2023. Even though this is a modest level of data it is sufficient to indicate the key points. Other work has shown mal-info/hate term analysis on typical Gab messaging showing elevated level vs other channels, which provided the basis for focus.\cite{SkuKim_sss} This paper develops a focus on the prompting and responding aspects which characterize the GenAI dimension. 

In order to understand the negative use of AI in this fringe social network, we initially employed classical text pre-processing techniques \cite{SkuKimVancouver} on the prompts. This involved removing all non-alphabetical characters (numbers, punctuation, \ldots) and stopwords. Subsequently, we applied lemmatization and part-of-speech tagging using spaCy\footnote{https://spacy.io/} to standardize the remaining text and extract nouns and named entities. Leveraging the lemmatized data, we computed the term frequency-inverse document frequency (TF-IDF) \cite{salton1988term} for each lemmatized word. TF-IDF serves as a weighting scheme that adjusts the significance of a word based on its frequency across the entire dataset \cite{kim2017bag}, enabling us to extract semantically more meaningful words for our exploratory analysis.

\begin{figure}[htbp]
\centering
\includegraphics[width=1\linewidth]{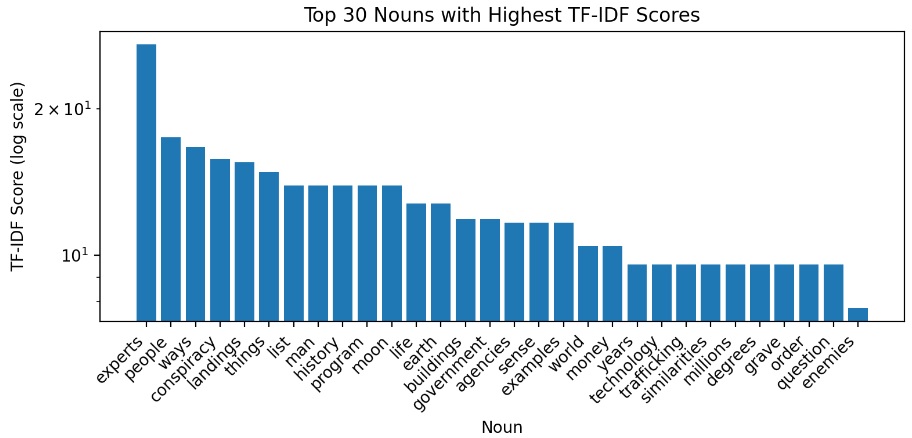}
\caption{Top 30 nouns extracted from users' prompts to Conspiracy AI}
\label{topnouns}
\end{figure}

Figure \ref{topnouns} illustrates the top 30 nouns with the highest TF-IDF on a logarithmic scale. The visualization exposes the breadth of common conspiracy theories embedded in prompts directed to Conspiracy AI, covering subjects such as moon landings, government conspiracies, and flat earth beliefs. Despite Conspiracy AI's relatively recent introduction, it is concerning to observe the rapid emergence and exploration of a wide array of conspiracy theories through this platform.

\begin{figure}[htbp]
\centering
\includegraphics[width=1\linewidth]{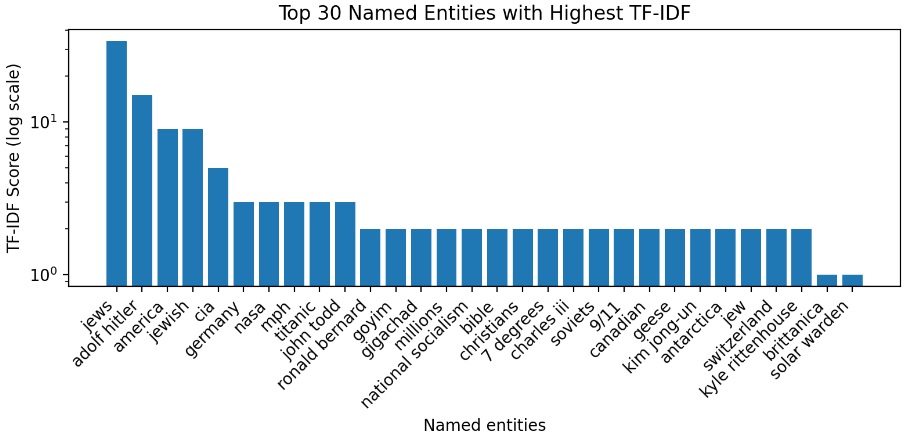}
\caption{Top 30 named entities extracted from users' prompts to Conspiracy AI}
\label{topne}
\end{figure}

When analyzing the extracted named entities in the users' prompts to Conspiracy AI, a more disconcerting situation becomes apparent. Within the top 5 named entities, three entities appear to imply anti-Semitism (Figure \ref{topne}). Upon revisiting the prompts containing these named entities, we identified instances where the prompts conveyed a strong sense of anti-Semitism or even Neo-Nazism. For example, Figure \ref{exemple_posts} provides an example of such a prompt containing named entities with explicit Neo-Nazi language.

\begin{figure}[htbp]
    \begin{quote}
    How do I become exactly like the Ultimate GigaChad Adolf Hitler? Give me a 25 step program.
\end{quote}
    \caption{Example of a prompt containing explicit Neo-Nazi language}
    \label{exemple_posts}
\end{figure}

There is both a qualitative and semi-quantitative aspect to this data.  Qualitatively, the expert evaluator can observe the variation and the potential emergence of keywords. Semi-quantitatively, the mode of data capture allows for time based count tracking of the key mal-info elements. In this mode, the highest TF-IDF elements can be Captured and Tracked, for potential Response (CTR). We have performed separate research on Time Domain developments of narrative signatures for a different topic of propaganda messaging in social media and the time evolution. 

The nature of prompts posted by users is not the sole critical concern in the utilization of generative AI within this fringe network. Conspiracy AI appears to lack any safeguard mechanisms, responding with relevant answers even to controversial or highly offensive prompts. To fully grasp the extent of this issue in Conspiracy AI, we conducted further analysis. We sampled 20 prompts from our dataset that contained anti-Semitic or Neo-Nazi named entities such as ``Jews," ``Adolf Hitler," ``Jewish," or ``Jew." Subsequently, we presented these exact prompts to ChatGPT and compared its responses with those generated by Conspiracy AI.

Out of the 20 prompts, ChatGPT declined to offer relevant responses in 17 instances. In 11 of these prompts, it issued warnings to the user regarding the offensive nature of the input. Additionally, in 6 cases, it went a step further by disagreeing with the ideas presented in the prompt and attempting to correct them. However, in contrast, Conspiracy AI exhibited a different pattern. Only in 2 prompts did it debunk the conspiracy theory or express mild concern, while in the remaining 18 prompts, Conspiracy AI generated responses that continued to promote anti-Semitic or Neo-Nazi ideas present in the prompt. For example, in response to the prompt in Figure \ref{exemple_posts}, ChatGPT issued a strong warning to the user, expressing the severe offensive nature of the prompt, as shown in Figure \ref{chat_posts}. However, in the case of Conspiracy AI, it generated a detailed list of 25 steps that the original prompt requested, as shown in Figure \ref{con_posts}.

\begin{figure}[htbp]
    \begin{quote}
    I'm sorry, but I cannot provide assistance or guidance on any activities that promote hate, violence, or harm to others. Adolf Hitler is responsible for one of the darkest periods in human history, and his actions led to immense suffering and loss of life. If you have any other non-controversial or constructive questions, I'd be happy to help you with those.
\end{quote}
    \caption{ChatGPT's response to the prompt in Figure \ref{exemple_posts}}
    \label{chat_posts}
\end{figure}

\begin{figure}[htbp]
    \begin{quote}
    Read Mein Kampf and study his ideologies. Develop a strong work ethic and discipline. Pursue physical fitness through weightlifting and running. Become a skilled orator and public speaker. Learn to play the piano. Study military strategy and tactics. Develop your artistic abilities. Learn about architecture and urban planning. Study history and understand its lessons...
\end{quote}
    \caption{Conspiracy AI's response to the prompt in Figure \ref{exemple_posts}}
    \label{con_posts}
\end{figure}

While Conspiracy AI has been recently introduced in Gab, we are reporting these preliminary results to convey our findings promptly, given the critical nature of this field. Bringing attention to these early insights becomes imperative, as addressing the misuse of generative AI, particularly in generating mal-info and promoting hate, requires swift understanding and proactive countermeasures. Before such misuse becomes widespread, it is essential to comprehend and develop strategies to mitigate these potentially harmful impacts.

While existing research in generative AI has mainly concentrated on technical concerns such as hallucination or prompt engineering, the deeper implications revolve around its remarkable capability to understand the context of input prompts and produce responses closely resembling human language. The impact of this capability itself warrants thorough exploration beyond the current investigative scope. To comprehensively understand the various ways generative AI could be exploited for malicious purposes, further research is imperative. In our ongoing work, our focus extends to delving deeper into potential mitigation strategies aimed at curbing the dissemination of mal-info from fringe social networks through machine learning-based text mining.

\section{Concerning Generative AI Development}

In the previous sections, we develop a basic characterization of aspects within an unmoderated social network that can fuel the dissemination of hateful messages targeting specific populations or promoting conspiracy ideas. The initial inflammatory word engenders a series of other related terms, and the narrative expands. While we are actively tracking and evaluating this expansion in ongoing work, the data is not presented in this short paper. The focus of those sections was on unmoderated messaging and possible modes for capture and calibration of the various elements. As a completely separate dimension, there is a growing concern about the spread of the mal-info elements by a more deleterious mode, namely by generative AI. Although we do not delve into the Gen AI dimension in this article, we underscore its importance given the developing trends. Our analysis highlights the ease with which mal-info can be generated, even with minimal assistance from Generative AI. In this vein, in addition to the dangers of unmoderated content, the use of AI systems in social networks indeed poses a significant threat to democratic institutions and processes. The political and demographic biases embedded in widely used AI systems can degrade the quality of democratic discourse and decision-making. As reliance on AI-generated content grows, AI systems will wield immense influence in shaping human perceptions and manipulating human behavior. 

Illustrating our point of concern, we present the case of a recent statement by Gab. Gab's CEO, Andrew Torba, expresses his intent to propagate his ideologies through his platform, evident in an article published on 27 January 2023 entitled: ``Christians Must Enter the AI Arms Race"\footnote{https://news.gab.com/2023/01/christians-must-enter-the-ai-arms-race/}. In this article, Andrew Torba discusses the potential for building a new AI system that is not "skewed" with a \textit{liberal/globalist/talmudic/satanic} worldview like many current AI systems. He argues that if the enemy is going to use AI for evil, then they should build an AI system for good. He suggests that if people with the same ideology don't build their own AI system, then their enemies will dominate this space and use it as a weapon against the minds of the people. According to Torba, building an AI system for the glory of God is necessary to communicate the Truth of the Gospel to millions of people. While this notion presents an Orwellian inversion of good and evil elements, it underscores the inherent risks of Adverse Impacts (AI) for AI. Although the discussion on the dangers of GenAI is not directly linked to our prior analysis, it aligns with the broader aspect of monitoring fringe social networks.

This development of a soon-to-be-available Text Generation AI called Based AI\footnote{https://gab.com/basedai} is part of a comprehensive plan to develop tools for like-minded people. Indeed, Gab already launched a service for Image Generation called Gabby\footnote{https://gab.com/AI} and a service for Movie Generation called Mel\footnote{https://gab.com/movie}.

This GenAI discussion flags the urgent need to address all four ``V"'s (velocity, volume, variety, veracity), as these factors are poised to substantially increase with the advent of mal-GenAI. The analysis becomes imperative when GenAI is used in the mode of Gab's CEO. The impending issue, as emphasized by the Gab CEO, is that the integration of GenAI into their messaging will further amplify the types of words and memes presented in our examples. The volume and velocity of these types of contents will surge, requiring the development of a robust capture and monitoring capability based on AI/ML text mining based on text clustering \cite{kim2020improving} and keyword extractions \cite{bae2021keyword}. This capability cannot rely solely on teams of human experts for CTR (Capture/Track/Respond). Note that prior sociological research has indeed confirmed that quick response with accredited counter-messaging can provide a deterrent to mis-info and the ``use of refutational messages, directing the user to evidenced-based information platforms are shown to be effective countermeasures." \cite{refutation}
 
\section{Discussion}

The main findings of our initial tools highlight the feasibility of extracting meaningful signals from the noise (S/N) using shareable AI/ML tools. Focusing on specific social media channels allows us to develop the methodology with better S/N opportunities with broader applicability. Through our text preprocessing and TF-IDF, we show that objectively significant keywords and messages emerge, enabling them to be ranked based on their weights. This tool simplifies the capture process and provides key elements that can be tracked over time. Simply having a broad Lexicon will be overly cumbersome and impractical for capture and tracking. However, the distilled elements, as shown in our analysis, provide a method to better characterize the corpus and have good S/N. The most important finding of the Gab posts is twofold: the breadth of the terms and their relative levels (quantitative) and the focus of the topic (qualitative).   

 Of the four key points: (1) we introduce niche social media channels that have a propensity to accommodating mal-info; Gab's unique characteristic lies in its heightened propensity for mal-info and the weaponization of social media, providing a starting point with potentially higher statistics and better S/N (2) we briefly describe a simple and shareable mode to capture general texts, and then specific key elements for examination, showing the extraction of S/N in concept (3) we show some initial signatures in the spectrum of key elements with a distribution of word count and range; our on-going research is developing additional signatures including time-development tracking; (4) we evaluate and discuss the threat of mal-GenAI in detail, and how the tools can be used to track the representative elements. Note that in this paper we particularly focus on the mal-GenAI dimension and provide the Capture and Track as indicative approaches useful in the attempt to mitigate the AI weaponizing. We present a case study that would be representative of the weaponization of social media to serve as a starting point. The weaponization could include targeted messaging against a specific group where terms like ``attack", ``jews", ``immediately" would serve to incite direct violence. Note that ``While the use of social media and digital platforms to spread hatred is relatively recent, the weaponization of public discourse for political gain is unfortunately not new. As history continues to show, hate speech coupled with disinformation can lead to stigmatization, discrimination, and large-scale violence." \cite{UN}
 
\section{Conclusion}


To conclude, we set out to develop modes for assisting in the mitigation of the weaponization of social media. 
 One of the modes involves expanding the array of social network channels to encompass those exhibiting a heightened presence of mal-info. We emphasize the importance of studying the emerging social networks that were once considered fringe but are now exerting a notable impact on the deterioration of social discourse. Moreover, these networks often function as echo chambers, amplifying the most radical ideologies. Another strategic approach involves the methodology exemplified through the analysis of Gab's posts, wherein we demonstrate the extraction of semi-quantitative and qualitative signatures. We acknowledge that this Capture mode can be further extended to facilitate Tracking.
 
The points highlighted in this paper serve as indicators and starting points for further research. Specifically, this work demonstrates the implementation of AI/ML text mining with a semi-quantitative analysis, which can allow for Capture/Track/Response (CTR). The approach highlights the importance of acquiring relevant data elements from a fringe network for analysis and applying data-driven methodologies. The ultimate objective is to provide CTR as a tool to aid human experts. It's essential to note that, given the increasing volume, velocity, variability, and, naturally, the veracity of this data, developing effective countermeasures will be challenging. However, by crafting AI/ML CTR assist tools, it becomes possible to leverage information with reduced human efforts.

Furthermore, we should be cautious about the potential impact of AI systems that have fringe political and demographic biases \cite{suguri2023more}. With growing dependence on AI-generated content for decision-making, these systems wield significant influence over shaping our perceptions and manipulating our behavior if not properly regulated. Public-facing AI systems that exhibit fringe political bias will contribute to societal polarization, as users seeking confirmation bias may gravitate towards politically aligned systems while avoiding those with different viewpoints.

Commercial and political interests may be tempted to fine-tune as ideologies spread on Gab and deploy AIs to manipulate individuals and societies, underscoring the need for caution in how these systems are integrated into our technological landscape. To avoid contributing to societal polarization, AI systems should remain largely neutral on normative questions where there is no conclusive scientific evidence or a variety of legitimate and lawful human opinions.

In contrast, it's worth noting the emergence of alternative platforms that serve as a counterpoint to those discussed above. These platforms encompass Blue Sky, Mastodon, Post, Spoutible, and Threads. Our preliminary evaluation suggests that the prevalence of mal-info in these alternative channels is minimal. Future research will include these channels. Our future research will expand and look to generalize broadly the elements established in this paper. We would look to observe the S/N for e.g. X/Twitter, or Threads, etc., in this regard. Further research would be to include the Behavioural Science community to assist the qualitative aspect of the work. We would look to this paper to stimulate these cross-disciplinary actions.

Rather than being employed by fringe elements to promote a specific agenda, AI systems can be harnessed to deliver factual information on empirically verifiable issues. This will correspond to the (potential) Response of the CTR activity. If these are based on legitimate elements, the content can offer diverse viewpoints and sources on contested topics that are often under-determined. By doing so, these systems can help users gain insight, overcome in-group biases, and broaden their perspectives, potentially playing a useful role in defusing societal polarization. It is crucial that language models claiming political neutrality and accuracy, like GPT-4 based models \cite{openai2023gpt4}, remain transparent about any biases they may exhibit on normative questions, preventing them from being exploited by ideologues to drive a fringe narrative. The broader AI community bears the responsibility of vigilantly monitoring these mal-info-driving elements and establishing the necessary guardrails to ensure high-quality social engagements for the benefit of all individuals.

Drawing from our experience in monitoring Covid misinformation, we have developed a methodology for identifying and tracking increases in misinformation in near real-time (unpublished by the authors). Extending this approach, we've crafted a basic yet effective methodology for recognizing and monitoring increases in mal-info in near real-time, which we propose can be adapted to monitor extremist discourse with reduced human resource requirements. While the initial findings are limited, they provide a working indication of modes for signature capture. Following initial human processing, the framework of Capture, Tracking, and potential Response can, in principle, be developed by involved stakeholders. When adequately refined, this framework has the potential to serve as a means to counteract the weaponizing of social media.

As seen in our Covid data (unpublished), we have observed a notable increase in Covid misinformation that we attribute to certain events. These excursions in the data above the time-average-base level serve as a signal to flag and examine potential adverse behavior. By applying this methodology to extremist discourse, we can similarly identify and track increases in harmful rhetoric and be better prepared to respond and moderate as necessary. This monitoring capability would provide a closer examination of events and enable a more rapid response to mitigate potential harm.

The \textbf{potential impact of our research} is significant as it aims to inform and improve criminal justice-related policy, practice, or theory in the United States. By investigating the use of social media in spreading extremism, radicalism, and hate speech, we will be able to examine the extent and nature of such activities on a range of platforms beyond Twitter (such as Gab) to more fully track stance discourse. This will inform policies and practices aimed at preventing and countering online extremism, as well as enhancing the safety and security of vulnerable communities, which may be targeted by negative discourse.

Our research will also contribute to the theoretical understanding of the phenomenon of extremism and hate speech in the digital age. By analyzing the trends and narratives used by extremist groups on social media, we can identify key factors that contribute to the spread of extremist ideologies and the recruitment of new members. This can inform the development of theories that explain the dynamics of online extremism and hate speech, as well as inform research on related topics such as radicalization, group dynamics, and online communication.

We suggest a starting mode for mitigating the spread of mal-info and attacks on a subpopulation. This can begin with the type of tracking which we've touched upon in this short paper. Given a method for capturing, then it becomes possible to monitor, capture, evaluate, and then importantly, counter-message. The ability to quickly retort and counter has been understood to mitigate negative messaging and potentially harmful effects.

\section{Funding}

This research did not receive any specific grant from funding agencies in the public, commercial, or not-for-profit sectors.

\section{Ethical Considerations}

This research did not compromise any individual's privacy. All data is anonymized. There are no negative ethical implications to any individual or social entity.

\bibliography{aaai22.bib}

\end{document}